\documentclass[fleqn,10pt]{wlscirep}
\title{Dynamical anchoring of distant Arrhythmia Sources by Fibrotic Regions via   Restructuring of the Activation Pattern}

\author[1,*]{Nele~Vandersickel}
\author[2]{Masaya~Watanabe}
\author[3]{Qian~Tao}
\author[4]{Jan~Fostier}
\author[2]{Katja~Zeppenfeld}
\author[1,2,5]{Alexander~V~Panfilov}

\affil[1]{Department of Physics and Astronomy,
  Ghent University, Belgium}
\affil[2]{Department of Cardiology, Leiden University Medical
  Center, Leiden, the Netherlands}
\affil[3]{Department of Radiology, Division of Image
  Processing, Leiden University Medical Centre, Leiden, the
  Netherlands}
\affil[4]{Department of Information Technology (INTEC), IDLab
  Ghent University --- imec, Ghent, Belgium}
\affil[5]{Ural Federal University, Ekaterinburg, Russia}

\affil[*]{nele.vandersickel@ugent.be}

\newcommand{\fig}[1]{Fig.\,\ref{fig:#1}}
\usepackage{float}


\begin{abstract}
Rotors are functional reentry sources identified in clinically relevant cardiac arrhythmias, such as ventricular and atrial fibrillation. Ablation targeting rotor sites has resulted in arrhythmia termination. Recent clinical, experimental and modelling studies demonstrate that rotors are often anchored around fibrotic scars or regions with increased fibrosis. However the mechanisms leading to abundance of rotors at these locations are not clear. The current study explores the hypothesis whether fibrotic scars just serve as anchoring sites for the rotors or whether there are other active processes which drive the rotors to these fibrotic regions. 
Rotors were induced at different distances from fibrotic scars of various sizes and degree of fibrosis. Simulations were performed in a 2D model of human ventricular tissue and in a patient-specific model of the left ventricle of a patient with remote myocardial infarction. 
In both the 2D and the patient-specific model we found that without fibrotic scars, the rotors were stable at the site of their initiation. However, in the presence of a scar, rotors were eventually dynamically anchored from large distances by the fibrotic scar via a process of dynamical reorganization of the excitation pattern. This process coalesces with a change from polymorphic to monomorphic ventricular tachycardia.
\end{abstract}
\begin{document}

\flushbottom
\maketitle
%
%
\thispagestyle{empty}

\section*{Introduction}
Many clinically relevant cardiac arrhythmias are organized by rotors. A rotor is an extension of the concept of a reentrant source of excitation into two or three dimensions with an area of functional block in its center, referred to as the core. Rapid and complex reentry arrhythmias such as atrial fibrillation (AF) and ventricular fibrillation (VF) are thought to be driven by single or multiple rotors. A clinical study by Narayan et al. \cite{Narayan2012} indicated that localized rotors were present in 68\% of cases of sustained AF. Rotors (phase singularities) were also found in VF induced by burst pacing in patients undergoing cardiac surgery \cite{Nash2006,Bradley2011} and in VF induced in patients undergoing ablation procedures for ventricular arrhythmias \cite{Krummen2014}. Intramural rotors were also reported in early phase of VF in the human Langendorff perfused hearts \cite{Nair2011,Jeyaratnam2011}. It was also demonstrated that in most cases rotors originate and stabilize in specific locations \cite{Krummen2014,ringenberg2014effects,Jeyaratnam2011, Nair2011,Haissaguerre2016}.\\
\\
A main mechanism of rotor stabilization at a particular site in cardiac tissue was proposed in the seminal paper from the group of Jalife \cite{Davidenko1992}. It was observed that rotors can anchor and rotate stable around small arteries or bands of connective tissue. Later, it was experimentally demonstrated that rotors in atrial fibrillation in a sheep heart can anchor in regions of large spatial gradients in wall thickness \cite{Yamazaki2012}. A recent study of AF in the right atrium of the explanted human heart \cite{Hansen2015} revealed that rotors were anchored by 3D micro-anatomic tracks formed by atrial pectinate muscles and characterized by increased interstitial fibrosis. The relation of fibrosis and anchoring in atrial fibrillation was also demonstrated in several other experimental and numerical studies  \cite{Skanes1998, Gonzales2014, Hansen2015, Haissaguerre2016,Zahid2016}. Initiation and anchoring of rotors in regions with increased intramural fibrosis and fibrotic scars was also observed in  ventricles \cite{arevalo2013tachycardia,ringenberg2014effects,Nair2011}.
One of the reasons for rotors to be present at the fibrotic scar locations  is that the rotors can be initiated at the scars (see e.g. \cite{arevalo2013tachycardia,ringenberg2014effects}) and therefore they can easily anchor at the surrounding scar tissue. However, rotors can also be generated due to different mechanisms, such as triggered activity\cite{antzelevitch2011overview}, heterogeneity in the refractory period  \cite{antzelevitch2011overview,weiss2005qu}, local neurotransmitter release  \cite{rosenshtraukh1989vagally,liu1997differing} etc. What will be the effect of the presence of the scar on rotors in that situation, do fibrotic areas (scars) actively affect rotor dynamics even if they are initially located at some distance from them? In view of the multiple observations on correlation of anchoring sites of the rotors with fibrotic tissue this question translates to the following: is this anchoring just a passive probabilistic process, or do fibrotic areas (scars) actively affect the rotor dynamics leading to this anchoring? Answering these questions in experimental and clinical research is challenging as it requires systematic reproducible studies of rotors in a controlled environment with various types of anchoring sites. Therefore alternative methods, such as realistic computer modeling of the anchoring phenomenon, which has been extremely helpful in prior studies, are of great interest. \\
\\
The aim of this study is therefore to investigate the processes leading to anchoring of rotors to fibrotic areas. Our hypothesis is that a fibrotic scar actively affects the rotor dynamics leading to its anchoring. To show that, we first performed a generic in-silico study on rotor dynamics in conditions where the rotor was initiated at different distances from fibrotic scars with different properties. We found that in most cases, scars actively affect the rotor dynamics via a dynamical reorganization of the excitation pattern leading to the anchoring of rotors. This turned out to be a robust process working for rotors located even at distances more than 10 cm from the scar region. We then confirmed this phenomenon in a patient-specific model of the left ventricle from a patient with remote myocardial infarction (MI) and compared the properties of this process with clinical ECG recordings obtained during induction of a ventricular arrhythmia.   
\section*{Results}
\subsection*{Dynamical anchoring of a distant Rotor in the 2D Model of Cardiac Tissue}
\begin{figure}[h]
  \centering
  \includegraphics[width=1.0\textwidth]{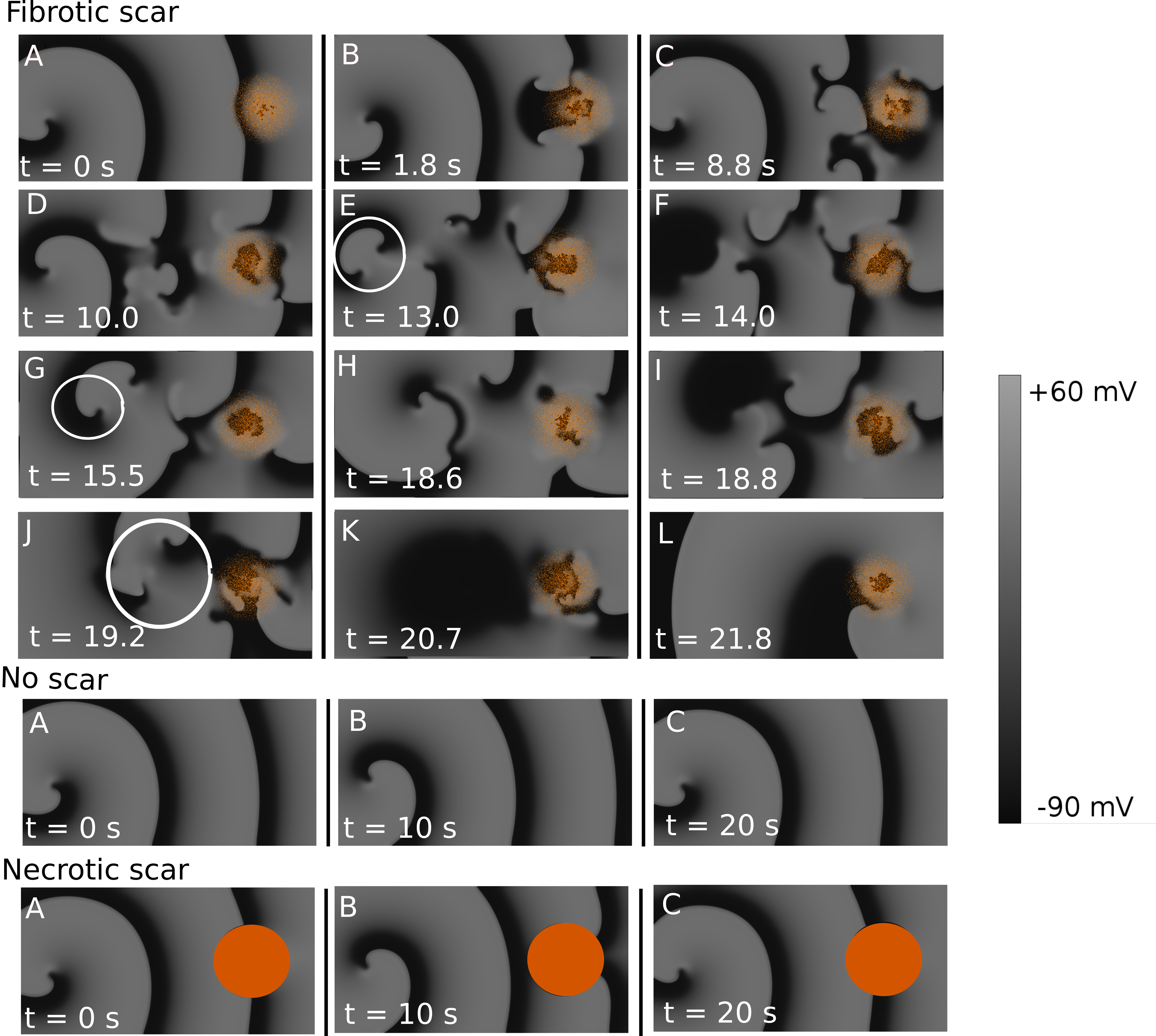}
  \caption{Dynamical anchoring of a distant rotor to a fibrotic region in the 2D model. Unexcitable fibrotic tissue is shown in orange and the transmembrane voltage is shown in shades of gray.\\
\textbf{Fibrotic scar}:\\
A: A rotor was initiated 15.5 cm away from the border of the fibrotic region. \\
B: The wavefront starts to break on the scar region forming secondary sources. \\
C-D: The secondary sources started to propagate towards the original rotor.\\ 
D: The secondary sources are reaching towards the initial rotor. \\ 
E: One secondary source will merge with the initial rotor, see the white circle. \\
F: The secondary source merged with the initial rotor, in the left side of the tissue there are no rotors anymore.\\
G: The main rotor is now indicated with the white circle, the process of rotor annihilation starts again.\\
H-I: A new rotor annihilation results again in a dominant rotor closer to the scar. \\
J-K: The dominant rotor merges again with a secondary source (white circle) which leads to restructuring of the activation pattern. \\
L: The rotor is now attached to the scar.\\
\textbf{No scar/Necrotic scar:}\\
A: A rotor was initiated at the left of the tissue, or 15.5 cm from the border the the necrotic scar.\\
B- C: This rotors remains stable during the simulation of 20 s.
  \label{fig:attr}}
\end{figure}

\begin{figure}[h]
  \centering
  \includegraphics[width=\textwidth]{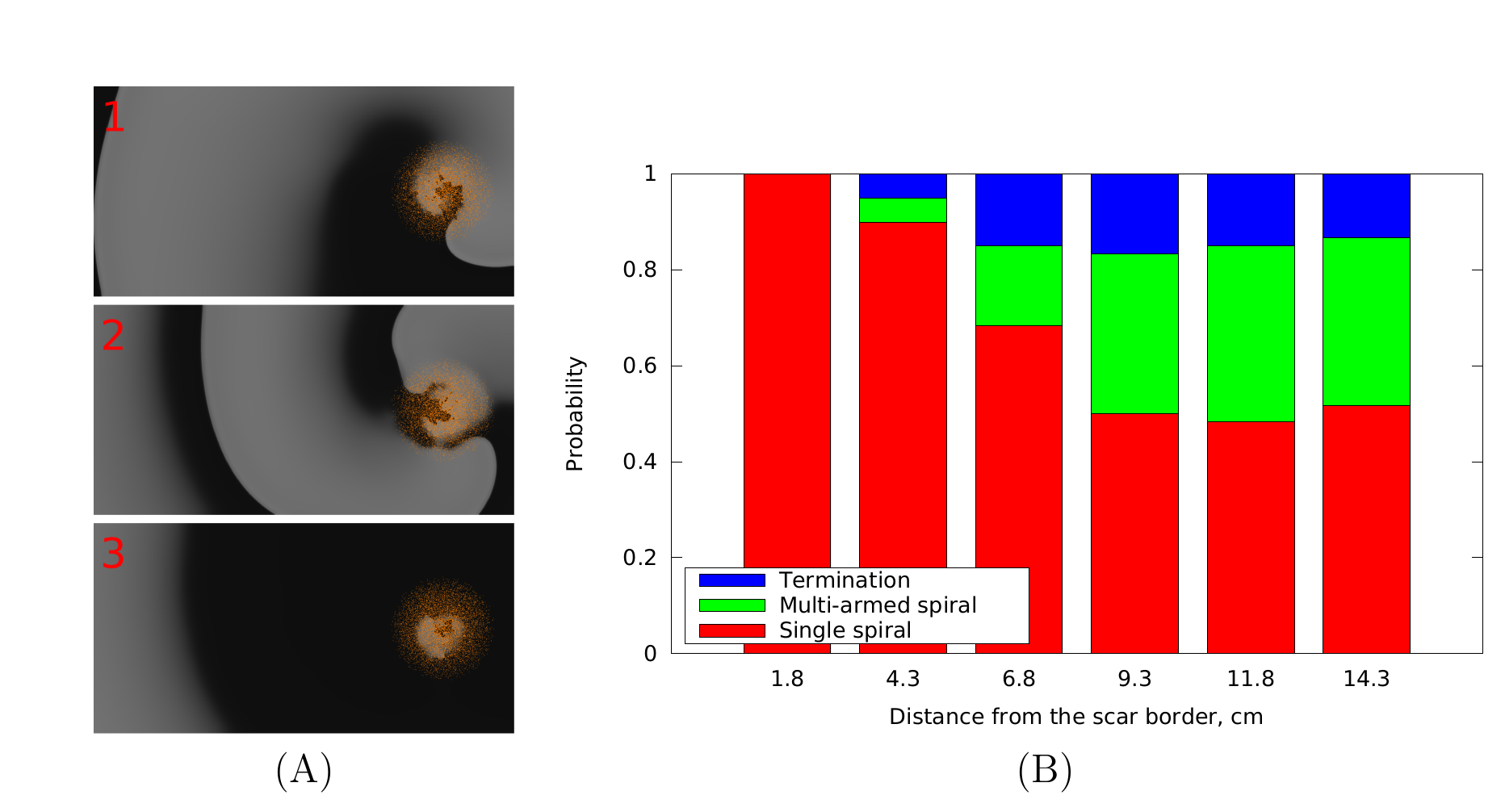} 
  \caption{panel A: Types of the resulting activation patterns: (1) a rotor rotating in the opposite direction than the original rotor, (2) a two-armed rotor and (3) termination of activity. Panel B: Type of the resulting activation pattern as a function of the distance between the rotor and the border of the scar. Red indicates: the resulting activation pattern is a rotor rotating either in the same direction as the original rotor or in the
    opposite direction. Green indicates: the resulting activation pattern is a rotor with two or three arms. Blue indicates: the electrical activity
    vanishes after dynamical restructuring of the activation pattern.  }
  \label{fig:types}
\end{figure}

Fibrotic scars can not only anchor the rotors but can dynamically anchor them from a large distance. In the first experiments we studied spiral wave dynamics with and without a fibrotic scar in a generic study. The diameter of the fibrotic region was 6.4 cm, based on the similar size of the scars from patients with documented and induced VT (see the methods section). The percentage of fibrosis changed linearly from 50\% at the center of the scar to 0\% at the scar boundary. We initiated a rotor at a distance of 15.5 cm from the scar (\fig{attr}, panel A) which had a period of 222 ms and studied its dynamics. Such distance is comparable with a distance between the RV apex catheter and a basal lateral LV scar. \\ 
\\
First, after several seconds the activation pattern became less regular and a few secondary wave breaks appeared at the fibrotic region (\fig{attr}, panel B). These irregularities started to propagate towards the tip of the initial rotor (\fig{attr}, panel C-D) creating a complex activation picture in between the scar and the initial rotor. Next, one of the secondary sources reached the tip of the original rotor (\fig{attr}, panel E). Then, this secondary source merged with the initial rotor (\fig{attr}, panel F), which resulted in a deceleration of the activation pattern and promoted a chain reaction of annihilation of all the secondary wavebreaks in the vicinity of the original rotor. At this moment, a secondary source located more closely to the scar dominated the picture (\fig{attr}, panel G). The whole process now started again (\fig{attr}, panels H-K), until finally only one source became the primary source anchored to the scar (\fig{attr}L) with a rotation period of 307 ms. For clarity, a movie of this process is provided as supplementary material.\\
\\
Note that this process occurs only if a scar with surrounding fibrotic zone was present. In the simulation entitled as  `No scar' in \fig{attr}, we show a control experiment  when the same initial conditions were used in tissue without a scar. In the panel entitled as  `Necrotic scar' in \fig{attr},  a simulation with only a compact region without the surrounding fibrotic tissue is shown. In both cases the rotor was stable and located at its initial position during the whole period of simulation. 
\noindent 
We refer to this complex dynamical process leading to anchoring of a distant rotor as dynamical anchoring. Although this process contains a phase of complex visually chaotic behaviour, overall it is extremely robust and reproducible in a very wide range of conditions. In the second series of simulation, the initial rotor was placed at different distances from the scar border, ranging from 1.8 to 14.3 cm, to define the possible outcomes, see \fig{types}. Here, in addition to a single anchored rotor shown in \fig{attr}H we could also obtain other final outcomes of dynamical anchoring: we obtained rotors rotating in the opposite direction (\fig{types}A, top), double armed anchored rotors which had 2 wavefronts rotating around the fibrotic regions (\fig{types}A, middle) or annihilation of the rotors (\fig{types}A, bottom, which show shows no wave around the scar), which normally occurred as a result of annihilation of a figure-eight-reentrant pattern. To summarize, we therefore had the following possible outcomes:
\begin{itemize}
\item Termination of activity
\item A rotor rotating either clockwise or counter-clockwise
\item A two- or three-armed rotor rotating either clockwise or
  counter-clockwise
\end{itemize}


\begin{figure}[h]
  \centering
  \includegraphics[width=\textwidth]{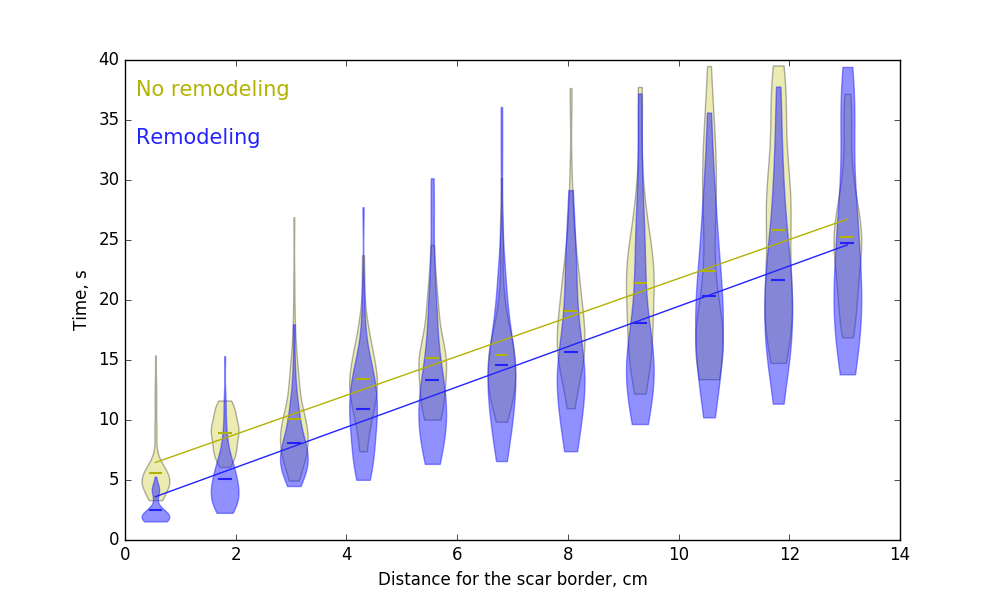}
  \caption{Violin plot of the dependency of time required for rotor anchoring 
  or termination on the initial distance between the rotor tip and the
    border of the fibrotic region in the 2D model.  Yellow indicates the model of    fibrosis whereby ionic remodeling was taking into account, while blue indicates that fibrosis is
    modeled only by introduction of small unexcitable obstacles.}
  \label{fig:distance}
\end{figure}

\noindent
\fig{types}, panel B presents the relative chance of the
mentioned activation patterns to occur depending on the distance
between the rotor and the border of the scar. We see,
indeed, that for smaller initial distances the resulting activation pattern is
always a single rotor rotating in the same direction.  With increasing distance, other anchoring  patterns are possible. If the distance was larger than about 9~cm, there is at least a 50\% chance to obtain either a multi-armed rotor or termination of activity.  Also note that such dynamical anchoring occurred from huge distances: we studied rotors located up to 14~cm from the scar. However, we observed that even for very large  distances such as 25~cm or more such dynamical anchoring (or termination of the activation pattern) was always possible, provided enough time was given.  \\
\\
We measured the time required for the anchoring of rotors as a function of the distance from the scar. For each distance, we performed about 60 computations using different seed values of the random number generator, both with and without taking ionic remodeling into account. The results of these simulations are shown in \fig{distance}. 
We see that the time needed for dynamical anchoring depends linearly on the distance between the border of the scar and the initial rotor. The blue and yellow lines correspond to the scar model with and without ionic remodeling, respectively. We interpret these results as follows. The anchoring time is  mainly determined by  the propagation of the chaotic regime towards the core of the original rotor and this process has a clear linear dependency. For distant rotors, propagation of this chaotic regime mainly occurs  outside the region of ionic remodelling, and thus both curves in Fig. 3 have the same slope.  
 However, in the presence of ionic remodelling, the APD in the scar region is prolonged. This  creates a heterogeneity and as a consequence  the initial breaks in the scar reqion are formed about $3.5$~s earlier in the scar model with remodeling compared with the scar model without remodeling.

\begin{figure}[h]
  \centering
  \includegraphics{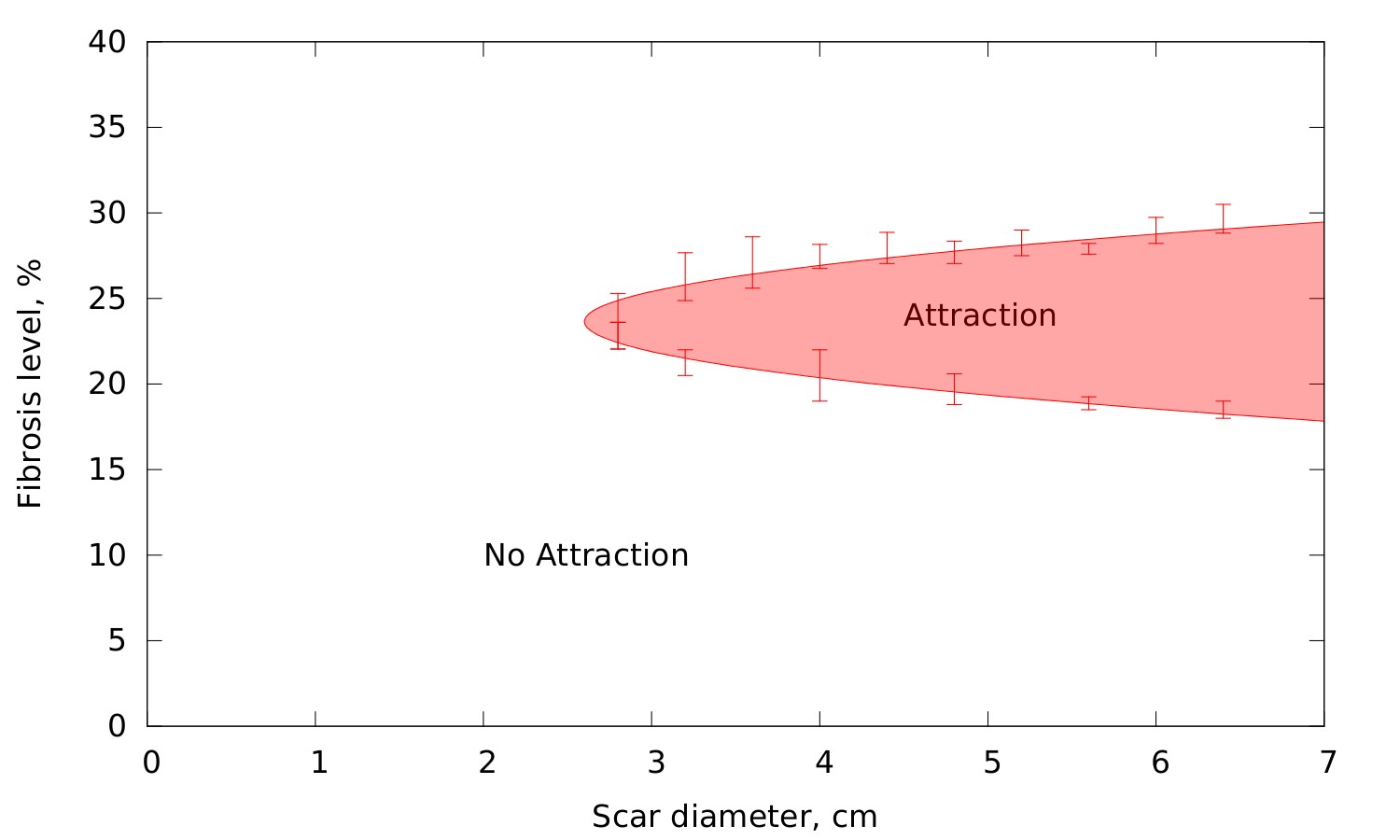}
  \caption{Phase diagram showing the region where the anchoring effect
    is present in two-parametric space: the size of the scar and the
    fibrosis level. The scar had a uniform fibrosis distribution. The
    mark ``Attraction'' corresponds to the region where the dynamical anchoring
    was obtained in more than 65\% of the cases. ``No Attraction''
    shows the region where the effect occurred in less than 65\% of
    the cases. The scar was located 5 cm from the scar.
  \label{fig:fibsize}}
\end{figure}\noindent
To identify some properties of the substrate necessary for the dynamical anchoring we varied the size and the level of fibrosis within the scar and studied if the dynamical anchoring was present. Due to the stochastic nature of the fibrosis layout we performed about 300 computations with different textures of the fibrosis for each given combination of the scar size and the fibrosis level. 
The results of this experiment are shown in \fig{fibsize}. Dynamical anchoring does not occur when the scar diameter was below $2.6$~cm, see \fig{fibsize}. For scars of such small size we observed the absence of both the breakup and dynamical anchoring. We explain this by the fact that if the initial separation of wavebreaks formed at the scar is small, the two secondary sources merge immediately, repairing the wave front shape and preventing formation of secondary sources \cite{Majumder2014}.\\
\\
Also, we see that this effect requires an intermediate level of fibrosis density. For small fibrosis levels no secondary breaks are formed. Also, for larger fibrosis levels, no breaks could be formed if the fibrosis level is larger than 41\% in our 2D model, as the tissue behaves like inexcitable scar. In this case, the scar effectively becomes a large obstacle that is incapable of breaking the waves of the original rotor \cite{Majumder2014}. Close to the threshold of 41\% we have also observed another interesting pattern when the breaks are formed inside the scar only and cannot exit to the surrounding tissue.\\ 
\\
Finally, note that \fig{fibsize} illustrates only a few factors important for the dynamical anchoring in a simple setup in an isotropic model of cardiac tissue.  The particular values of the fibrosis level and the size of the scar can also depend on anisotropy, the texture of the fibrosis and its possible heterogeneous distribution.

\subsection*{Dynamical anchoring  in the Patient Specific Model of the Left Ventricle}
\begin{figure}[h]
  \centering
  \includegraphics[width=1.0\textwidth]{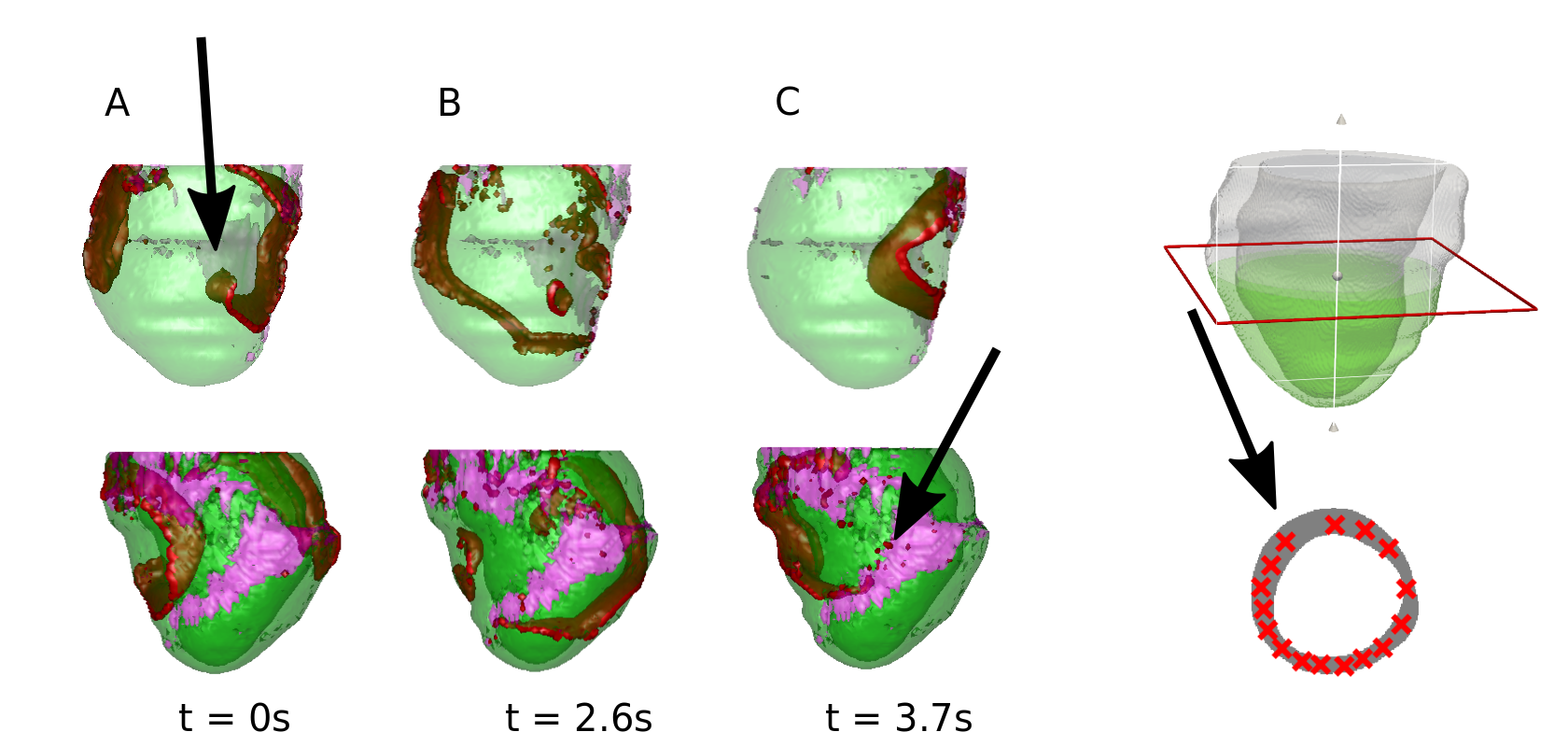}
  \caption{Effect of the dynamical anchoring of a rotor for a patient-specific
    model of the left ventricle after infero-lateral MI. Red color corresponds to high
    transmembrane voltage, pink color shows the scar. The top row shows the left ventricle in an modified anterior view (where the spiral was initiated) and the second row in a modified posterior view (location of scar). A: A rotor is initiated 5~cm away from the scar region. B: The breakups form
    making the activation pattern less regular (2.6~s after the
    initialization) C: The rotor gets anchored to the scar and rotates
    around it persistently (3.7~s after the initialization). D: The 16 different locations of the initial core of the rotor in a slice at 4 cm from the apex. In the bottom row, the simulations are shown where there is no scar present, and the spiral core stays around the same location. For a movie of the scar simulation, see the supplementary material.}
  \label{fig:attr3d}
\end{figure}

\begin{figure}[h]
  \centering
  \includegraphics{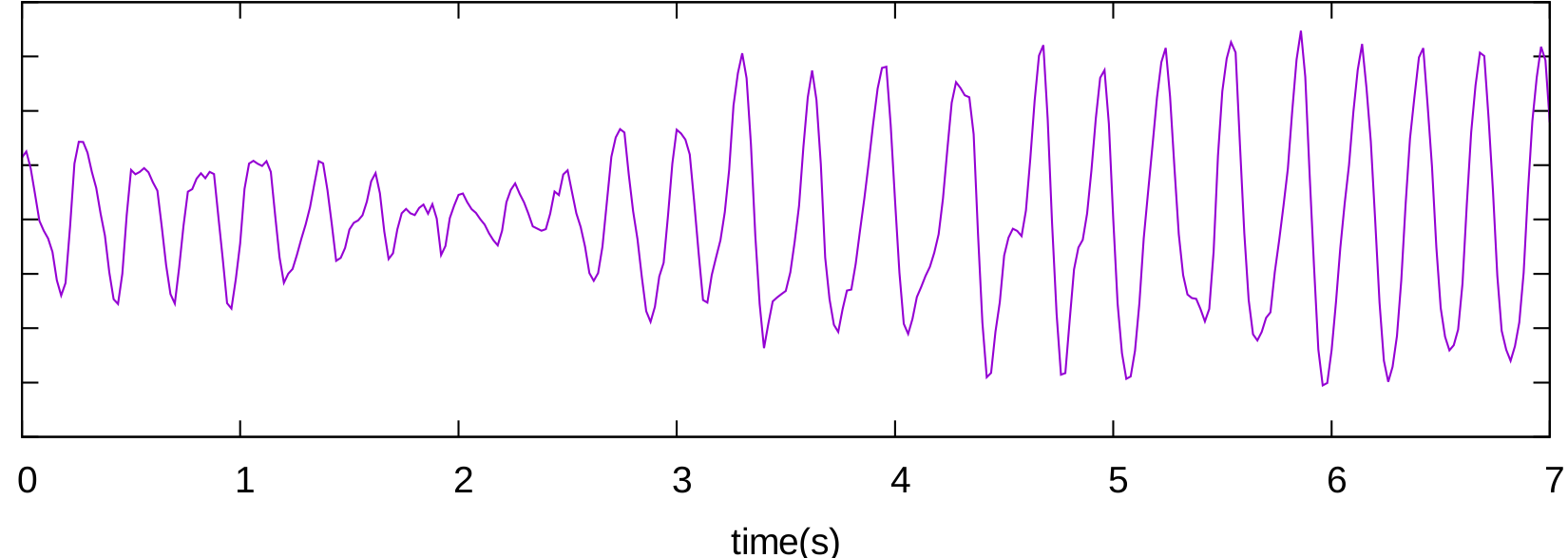}
  \caption{A representative example of a computed ECG for the anchoring effect in the 3D model is shown. The ECG is displayed in purple, without showing the units as only relative units are important. Anchoring is observed approximately at the end of the third second. The coupling interval increased after that time.
}
  \label{fig:modecg}
\end{figure}

To verify that the dynamical anchoring takes place in a more realistic geometry, we developed and investigated this effect in a patient-specific model of the human left ventricle, see the method section for details. 
The scar in this data set has a complex geometry with several compact regions with size around  5-7 cm in which  the percentage of fibrosis changes gradually from 0\% to 41\% at the core of the scar based on the imaging data, see methods section. The remodeling of ionic channels at the whole scar region was also included to the model (including borderzone as described the Fibrosis Model in the method section). We studied the phenomenon of dynamical anchoring for 16 different locations of cores of the rotor randomly distributed in a slice of the heart at about 4 cm from the apex  (see \fig{attr3d}).\\
\\
Cardiac anisotropy was generated by a rule-based approach described in details in the Methods
section. Of the 16 initial locations, shown in \fig{attr3d}, there was dynamical anchoring to the fibrotic tissue in all cases. After the anchoring, in 4 cases the rotor annihilated.\\
\\
A representative example of our 3D simulations is shown in
\fig{attr3d}.  We followed the same protocol as for the 2D
simulations.  The top 2 rows the modified anterior view and the modified posterior view in the case the scar was present. In column A we see the original location of the spiral core (5 cm from the scar) indicated with the black arrow in anterior view. In column B, breaks are formed due to the scar tissue, and the secondary source started to appear. After 3.7 s , the spiral is anchored around the scar, indicated with the black arrow in the posterior view, and persistently rotated around it. In the bottom row, we show the same simulation but the scar was not taken into account. In this case, the spiral does not radically change its original location (slight movement, see the black arrows).\\
\\
To evaluate if this effect can potentially
be registered in clinical practice we computed the ECG for our 3D simulations. The ECG that corresponds to the example in \fig{attr3d} is shown in \fig{modecg}. During the first three seconds the ECG shows QRS complexes varying in amplitude and shape and then more uniform beat-to-beat QRS morphology with a larger amplitude. This change in morphology is associated with anchoring  of
the rotor which occurs around three seconds after the start of the simulation. The initial irregularity is due to the presence of the secondary sources that have a slightly higher period than the original rotor.  After the rotor is anchored, the pattern becomes relatively stable which corresponds to a regular ECG morphology. Additional ECGs for the cases of termination of the arrhythmia and anchoring  are shown in supplementary Fig. 2. For the anchoring dynamics we see similar changes in the ECG morphology as in \fig{modecg}.\\
\\
The dynamical anchoring is accompanied by an increase of the cycle length (247 $\pm$ 16 ms versus 295 $\pm$ 30 ms). The reason for this effect is that the
rotation of the rotor around an obstacle --anatomical reentry-- is usually slower
than the rotation of the rotor around its own tip—functional
reentry. 
\noindent

\noindent

\subsection*{Clinical ECGs related to the dynamical anchoring process}%
\begin{figure}[h]
  \centering
  \includegraphics[width=\textwidth]{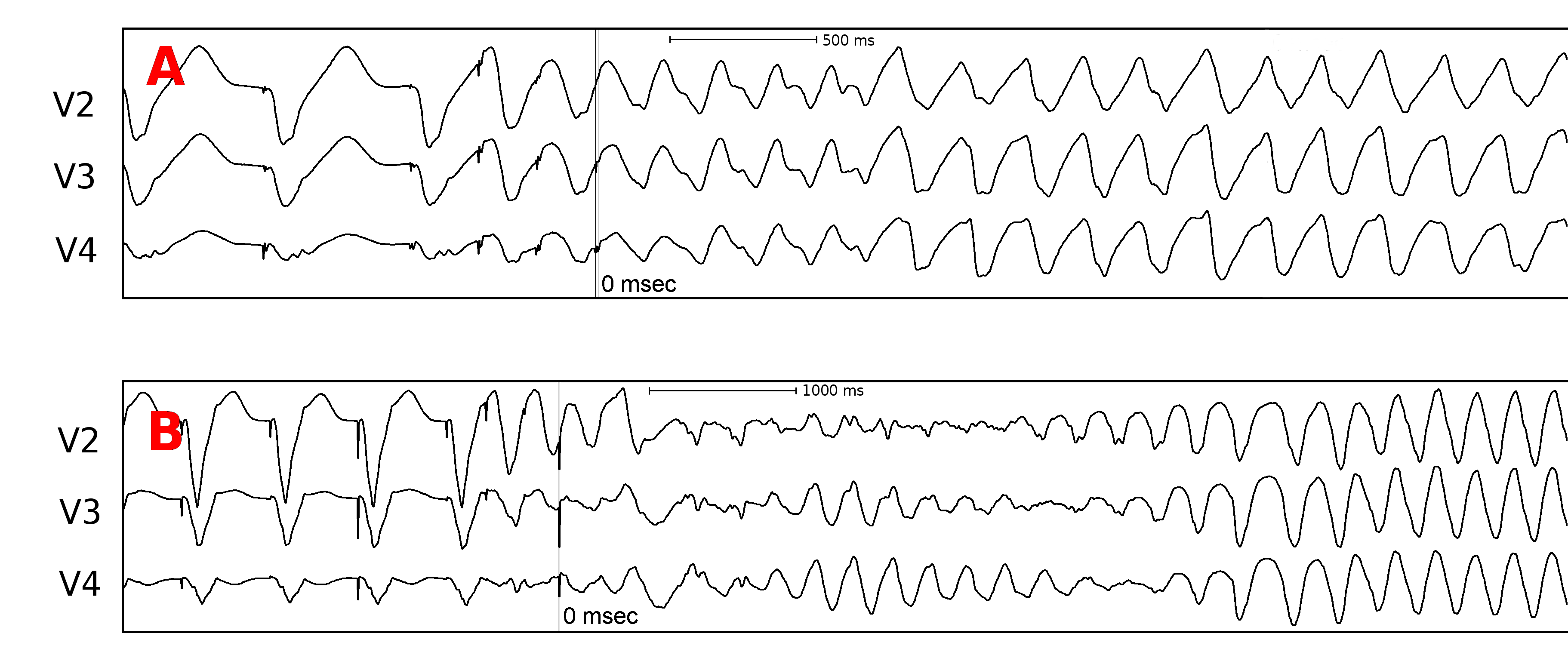}
  \caption{Precordial ECG leads V2, V3, and V4 recorded during induction of a ventricular tachycardia in two patients with scars in the left ventricle. First beats are paced beats; the mark “0 msec” indicates the last extrastimuli during programmed ventricular stimulation.}
  \label{fig:cliecg}
\end{figure}

In the previous section, we showed that the described results on dynamical anchoring in an anatomical model of the LV of patients with post infarct scars correspond to the observations on ECGs during initiation of a ventricular arrhythmia. After initiation, in 18 out of 30 patients (60\%) a time dependent change of QRS morphology was observed. Precordial ECG leads V2, V3 and V4 from two patients are depicted in \fig{cliecg}. For both patients the QRS morphology following the extra stimuli gradually changed, but the degree of changes here was different. In patient A, this morphological change is small and both parts of the ECG may be interpreted as a transition from one to another MVT morphology. However for patient B the transition from PVT to MVT is more apparent. In the other 16 cases we observed different variations between the 2 cases presented in \fig{cliecg}. Supplementary Fig. 3 shows examples of ECGs of 4 other patients. Here, in patients 1 and 2 we see substantial variations in the QRS complexes after the arrhythmia initiation and subsequently a transformation to MVT. The recording in patient 3 is less polymorphic and in patient 4 we observe an apparent shift of the ECG from one morphology to another. It may occur, for example, if due to underlying tissue heterogeneity additional sources of excitation are formed by the initial source. Overall the morphology with clear change from PVT to MVT was observed in 5/18 or 29\% of the cases. These different degrees of variation in QRS morphology may be due to many reasons, namely the proximity of the created source of arrhythmia to the anchoring region, the underlying degree of heterogeneity and fibrosis at the place of rotor initiation, complex shape of scar, etc.\\
\\
Although this finding is not a proof, it supports that the anchoring phenomenon may occur in clinical settings and serve as a possible mechanism of fast VT induced by programmed stimulation.

\noindent
\section*{Discussion}
In this study we investigated the dynamics of arrhythmia sources –rotors– in the presence of fibrotic regions using mathematical modeling. We showed that fibrotic scars not only anchor but also induce  secondary sources and  dynamical  competition of these sources normally results their annihilation. 
As a result if one just compares the initial excitation pattern in Fig.1A and final excitation pattern in Fig.1L it looks as if a distant spiral wave was attracted and anchored to the scar. However, this is not the case and the anchored spiral here is a result of normal anchoring and competition of secondary sources we we call dynamical anchoring.  This process is different from the usual drift or meandering of rotors where the rotor gradually changes its spatial position. In dynamical anchoring, the break formation happens in the fibrotic scar region, then it spreads to the original rotor and merges with this rotor tip and reorganizes the excitation pattern. This process repeats itself until a rotor is anchored around the fibrotic scar region. Dynamical anchoring may explain the organization from fast polymorphic to monomorphic VT, also accompanied by prolongation in CL, observed in some patients during re-induction after radio frequency catheter ablation of post-infarct scar related VT.\\ 
\\
In our simulations the dynamics of rotors in 2D tissue was stable and for given parameter values they do not drift or meander. This type of dynamics was frequently observed in cardiac monolayers \cite{Zlochiver2008,bingen2012prolongation} which can be considered as a simplified experimental model for cardiac tissue.  We expect that more complex rotor dynamics would not affect our main 2D results, as drift or meandering will potentate the disappearance of the initial rotor and thus promote anchoring of the secondary wavebreaks. In our 3D simulations in an anatomical model of the heart the dynamics of rotors is not stationary and shows the ECG of a polymorphic VT ( Fig.6). \\
\\
Although the dynamical anchoring reported in this paper will always bring the rotor to the scar region, the precise location of the rotor inside this region was not studied here. This question requires additional investigation, which is currently performed in our research group. The first preliminary results indicate that in case of a scar with a complex structure multiple anchoring sites are possible. Their location depends on several factors, such as the location of the initial rotor, the presence and the extent of the ionic remodelling. Identification of the specific features of the anchoring region and its delineation is of great importance as it may have implications for the treatment of the related arrhythmias.\\
\\
Similar processes can not only occur at fibrotic scars, but also at ionic heterogeneities. In \cite{Defauw2014}, it has been shown that rotors can be attracted by ionic heterogeneities of realistic size and shape, similar to those measured in the ventricles of the human heart \cite{Glukhov2010}. These ionic heterogeneities had a prolonged APD and also caused wavebreaks, creating a similar dynamical process as described in \fig{attr}. In this study however, we demonstrated that structural heterogeneity is sufficient to trigger this type of dynamical anchoring. \\
\\
The dynamical anchoring combines several processes: generation of new breaks at the scar, spread of breaks toward the original rotor, rotor  disappearance and anchoring or one of the wavebreaks at the scar. The mechanisms of the formation of new wavebreaks at the scar has been studied in several papers \cite{arevalo2013tachycardia,Majumder2014,cabo1996vortex} and can occur due to ionic heterogeneity in the scar region or due to electrotonic effects \cite{cabo1996vortex}. However the process of spread of breaks toward the original rotors is a new type of dynamics and the mechanism of this phenomenon remains to be studied. To some extent it is similar to the global alternans instability  reported in \cite{Vandersickel2016a}. Indeed in \cite{Vandersickel2016a} it was shown that an area of 1:2 propagation block can extend itself towards the original spiral wave and  is related to  the restitution properties of cardiac tissue. Although in our case we do not have a clear 1:2  block, wave propagation in the presence of breaks is disturbed resulting in spatially heterogeneous change of diastolic interval which  via the restitution effects can result in breakup extension. This phenomenon needs to be further studied as it may provide new ways for controlling rotor anchoring processes and therefore can affect the dynamics of a cardiac arrhythmia.\\
\\
In this paper we used the standard way of representation of fibrosis by placement of electrically uncoupled unexcitable nodes with no-flux boundary conditions. Although such representation is a simplification based on the absence of detailed 3D data, it does reproduce the main physiological effects observed in fibrotic tissue, such as formation of wavebreaks, fractionated electrograms, etc \cite{Vigmond2016}. The dynamical anchoring reported in this paper occurs as a result of the restructuring of the activation pattern and relies only on these basic properties of the fibrotic scar as the ability to generate wavebreaks and the ability to anchor rotors, which is reproduced by this representation. In addition, for each data point we performed simulations with at least 60 different textures. Therefore, we expect that the effect observed in our paper is general and should exist for any possible representation of the fibrosis. The specific conditions, e.g. the size and degree of fibrosis necessary for dynamical anchoring may depend on the detailed  fibrosis structure and it would be interesting to perform simulations with detailed experimentally based 3D structures of the fibrotic scars, when they become available. \\
\\
In the current paper we considered fibroblasts as inexcitable cells which are electrically uncoupled from the myocytes. It was suggested that formation of a scar can potentially enhance  fibroblast-myocyte coupling \cite{Vasquez2010} and such coupling has been recently measured in the scar from epicardial cryoinjury in the mice heart \cite{Quinn2016}. It will be interesting to study if fibroblast-myocyte coupling can influence the anchoring of rotors studied in this paper or affect the stability of anchored rotors. The effect of such coupling can be biphasic and can depend on the coupling strength and the resting potential of the fibroblasts \cite{Zlochiver2008,Sridhar2017}.\\
\\
In our simulations the dynamics of rotors in 2D tissue was stable and for given parameter values they do not drift or meander. Although this type of dynamics was observed in cardiac monolayers \cite{zlochiver2008electrotonic,bingen2012prolongation}, the size of the myocardial cell in cultured monolayer is usually smaller than myocytes in myocardium, and they do not have a cylindrical form. Moreover, the gap junctions in cell cultures are usually found circumferentially, whereas in vivo gap junctions are found mostly at intercalated disks \cite{angst1997dissociated}. As these differences may affect rotors, i.e. cause more complex dynamics, it may also affect the results of 2D studies.\\
\\
In-silico studies \cite{McDowell2015,Zahid2016} on modeling fibrosis in AF demonstrated that in AF, the reentrant activities co-located at the borders of fibrotic regions. Multiple experimental and clinical studies also showed co-localization of rotors and fibrosis \cite{Skanes1998,ringenberg2014effects, Gonzales2014, Hansen2015, Haissaguerre2016,Zahid2016,Nair2011}. Our results suggest that the possible mechanisms for the fact that such patterns are so abundant is due to dynamical anchoring. \\
\\
One of the possible effects by which fibrotic scars can influence rotor dynamics is the electrotonic influence from the fibrotic scar region. Indeed electrotonic effects are well known in cardiac electrophysiology and can strongly affect the heterogeneity of cardiac tissue and the susceptibility to arrhythmias \cite{billman2015cardiac}. It was also estimated in many models of cardiac cells that the spatial length of the electrotonic effects is of the order of 0.5-1cm \cite{sampson2005electrotonic,defauw2013action}. In our case we see dynamical anchoring for spirals located as far as 10-12 cm, which is far beyond these values. In addition, we also observe the same effect in case if we do not have ionic remodelling in the scar region (Fig.3), and thus in that case AP of all cells are the same (upto some possible boundary effects). Therefore, we think that the electrotonic influence from the scar is unlikely to be a main determinant of the dynamic anchoring. However, a heterogeneity around the scar has some effect on the anchoring process which is also can be seen in Fig.3

\textbf{Limitation}\\
In this paper we considered the case of a pre-existing rotor and focused on its interaction with the fibrotic scar. The formation of a rotor can occur via multiple possible mechanisms (e.g. \cite{arevalo2013tachycardia,antzelevitch2011overview,weiss2005qu,rosenshtraukh1989vagally}) which we did not take into account. This is because we wanted to address the process at a stage common for all mechanisms. This assumption is idealized, and it would be more natural to consider the complete sequence of transition from the sinus rhythm to rotor formation and then to its interaction with the scar. This is because such interaction process can potentially influence the process of dynamic anchoring. However, such interactions will add additional levels of complexity to the problem on top of the effects studied in our paper and may be specific for each particular mechanism. Therefore we decided to focus on this later stage of an already existing rotor, which is common for all these mechanisms. This is a limitation of our approach and it should be addressed in subsequent studies for each of the mechanisms of generation of the initial rotor..\\
\\
We have studied the dependency of dynamical anchoring to the scar on size and fibrotic content of the scar in a simplified situation: the scar was of circular shape, the fibrosis was modelled as diffuse fibrosis and had a constant level within the entire scar (Fig. 4). It would be interesting to extend such studies and consider different shapes of the scar and study the possible effects of this shape on the anchoring. Additionally, we could consider a non-homogeneous distribution of fibrosis inside the scar and study it with and without of ionic heterogeneity. Furthermore, it would be interesting to study the possible effects of different texture of fibrosis like interstitial and patchy patterns on the dynamical anchoring. Also, in this paper we have always considered an initial rotor rotating in a homogeneous part of the tissue. It would be interesting to study more realistic tissue setups, where fibrosis is not only present around the scar but also at distant locations where the rotor is present. We plan to address these shortcomings in subsequent research. Our 2D simulations are performed for rotors which have stationary rotation. As stationary rotation of free non-anchored rotors is unlikely to occur in the whole heart, these results account for an idealized situation and can change if the rotor rotation is not stationary. However, as we discussed only a qualitative relation between the data we think that our ‘general’   interpretation is acceptable. Clinical ECG recordings used in our paper were taken after ablation. It would be interesting to compare pre-ablation ECGs showing arrhythmia dynamics in patients with collected LGE MRI images. Unfortunately such clinical information was not available to us.

\section*{Methods}
\subsection*{Magnetic Resonance Imaging}
Our anatomical model is based on an individual heart of a post-MI patient  reconstructed from late gadolinium enhanced (LGE) magnetic resonance imaging (MRI) was described in detail previously \cite{Piers2014}. Briefly, a 1.5T Gyroscan ACS-NT/Intera MR system (Philips Medical Systems, Best, the Netherlands) system was used with standardized cardiac MR imaging protocol. The contrast –gadolinium (Magnevist, Schering, Berlin, Germany) (0.15 mmol/kg)– was injected 15 min before acquisition of the LGE sequences. Images were acquired with 24 levels in short-axis view after 600—700 ms of the R-wave on the ECG within 1 or 2 breath holds. The in-plane image resolution is 1 mm and through-plane image resolution is 5 mm.  Segmentation of the contours for the endocardium and the epicardium was performed semi-automatically on the short-axis views using the MASS software (Research version 2014, Leiden University Medical Centre, Leiden, the Netherlands). \\
\\
The myocardial scar was identified based on signal intensity (SI) values using a validated algorithm as described by Roes et al. \cite{Roes2009}. In accordance with the algorithm the core necrotic scar is defined as a region with SI $>$50\% of the maximal SI. Regions with lower SI values were considered as border zone areas. In these regions, we assigned the fibrosis percentage as normalized values of the SI as in Vigmond et al.  \cite{Vigmond2016}.  In the current paper, fibrosis was introduced by generating a random number between 0 and 1 for each grid point and if the random number was less than the normalized SI at the corresponding pixel the grid point was considered as fibroblast.\\
\\
Currently there is no consensus on how the SI values should be used  for clinical assessment of myocardial fibrosis and various methods have been reported to produce significantly different results \cite{Mewton2011}.  However, the method from Vigmond et al. properly describes the location of the necrotic scar region in our model as for the fibrosis percentage of more than 41-50\% we observe a  complete block of propagation inside the scar.

\subsection*{Electrophysiology Model}
The approach and the 2D model was described in detail in \cite{Tusscher2007a,Kazbanov2016}. Briefly, for ventricular cardiomyocyte we used the ten Tusscher and Panfilov (TP06) model \cite{tenTusscher2004,tenTusscher2006}, and the cardiac tissue was modeled as a rectangular grid of $1024\times512$ nodes. Each node represented a cell that occupied an area of $250\times250$~$\mu$m$^2$. The equations for the transmembrane voltage are given by
\begin{equation}
 C_m \frac{\mathrm{d}V_{ik}}{\mathrm{d}t} =
   \sum_{\alpha,\beta \in \{-1,+1\}} \eta_{ik}^{\alpha\beta}
   g_{\text{gap}} \left(V_{i+\alpha,k+\beta} - V_{ik}\right) -
   I_{\text{ion}}(V_{ik}, \ldots),
   \label{eq:tnnp}
\end{equation}
where $V_{ik}$ is the transmembrane voltage at the $(i,k)$
computational node, $C_m$ is membrane capacitance, $g_\text{gap}$ is the conductance of the gap junctions connecting two neighboring
myocytes, $I_\text{ion}$ is the sum of all ionic currents and
$\eta_{ik}^{\alpha\beta}$ is the connectivity tensor whose elements
are either one or zero depending on whether neighboring cells are coupled or not. A similar system of differential equations was used for the 3D
computations where instead of the 2D connectivity tensor
$\eta_{ik}^{\alpha\beta}$ we used a 3D weights tensor
$w_{ijk}^{\alpha\beta\gamma}$ whose elements were in between 0 and 1,
depending both on coupling of the neighbor cells and anisotropy due to
fiber orientation.  Each node in the 3D model represented a cell of the size
of $250\times250\times250$~$\mu$m$^3$.

\subsection*{Fibrosis Model}
Fibrosis was modeled by the introduction of electrically uncoupled unexcitable nodes \cite{TenTusscher2007}. The local percentage of fibrosis determined the probability for a node of the computational grid to become an unexcitable obstacle, meaning that for high percentages of fibrosis, there is a high chance for a node to be unexcitable. As  previous research has demonstrated that LGE-MRI enhancement correlates with regions of fibrosis identified by histological examination \cite{McGann2013} we linearly interpolated the SI into the percentage of fibrosis.\\
\\ 
 In addition, the effect of ionic remodeling in fibrotic regions was taken into account for several results of the paper  \cite{Pinto1999, Nattel2007}. To describe ionic remodeling we decreased the conductance of $I_\text{Na}$, $I_\text{Kr}$, and $I_\text{Ks}$ and  depending on local fibrosis level as:
\begin{eqnarray}
  G_\text{Na} &=& \left(1 - 1.55 \frac{f}{100\%} \right) G_\text{Na}^0, \\
  G_\text{Kr} &=& \left(1 - 1.75 \frac{f}{100\%} \right) G_\text{Kr}^0, \\
  G_\text{Ks} &=& \left(1 - 2 \frac{f}{100\%} \right) G_\text{Ks}^0,
\end{eqnarray}
where $G_X$ is the peak conductance of $I_X$ ionic current, $G_X^0$ is the peak conductance of the current in the absence of remodeling, and $f$ is the local fibrosis level in percent.  These formulas yield a reduction of 62\% for $I_\text{Na}$, of 70\% for $I_\text{Kr}$, and of 80\% for $I_\text{Ks}$ if the local fibrosis $f$ is 40\%.  These values of reduction are, therefore, in agreement with the values published in
\cite{Pu1997, Jiang2000}. \\
\\
The normal conduction velocity at CL 1000 ms is 72 cm/s (CL 1000 ms). However, as the compact scar is surrounded by fibrotic tissue, the velocity of propagation in that region gradually decreases with the increase in the fibrosis percentage. For example for fibrosis of 30\% the velocity decreases to 48 cm/s (CL 1000 ms). 

\subsection*{Model of the Human Left Ventricle}
The geometry and extent of fibrosis in the human left ventricles were determined using the LGE MRI data. The normalized signal intensity was used to determine the density of local fibrosis. The fiber orientation is presented in detail in the supplementary material.

\subsection*{Numerical Methods and Implementation}
The model for cardiac tissue was solved by the forward Euler
integration scheme with a time step of 0.02~ms. The numerical
solver was implemented using the CUDA toolkit
for performing the computations on graphical processing units.
Simulations were performed on a GeForce GTX Titan Black graphics card using single precision calculations.  \\
\\
The eikonal equations for anisotropy generation were solved
by the fast marching Sethian's method \cite{Sethian1996}.  The eikonal
solver and the 3D model generation pipeline were implemented in
the OCaml programming language.

\subsection*{Rotor initiation, termination and anchoring and pseudo-ECG computation}
Rotors were initiated by an S1S2 protocol, as shown in the supplementary Fig. 1. Similarly, in the whole heart simulations, spiral waves (or scroll waves) were created by an S1S2 protocol. \\
\\
For the compact scar geometry used in our simulations the rotation of the spiral wave was stationary, the period of rotation of the anchored rotor was always more than 280 ms, while the period of the spiral wave was close to 220 msec. Therefore, we determined anchoring as follows: if the period of the excitation pattern was larger than 280 ms over a time interval of 320 ms we classified the excitation as anchored.  When the type of anchoring pattern was important (single or multi-armed spiral wave) we determined it visually. If in all points of the tissue, the voltage was below -20 mV, the pattern was classified as terminated.  We applied the classification algorithm at t = 40 s in the simulation. \\
\\
In the whole heart, the pseudo ECGs were calculated by assuming an infinite volume conductor and calculating the dipole source density of the membrane potential $V_m$ in all voxel points of the ventricular myocardium, using the following equation \cite{plonsey1989}
\begin{equation} \label{eq_ecg}
 ECG( t) = \int \frac{ ( \vec r , D(\vec{r}) \vec \nabla V (t))}{ |\vec r|^3} d^3 r
\end{equation}
whereby $D$ is the diffusion tensor, V is the voltage, and $\vec r$ is the vector from each point of the tissue to the recording electrode. The recording electrode was placed 10 cm from the center of the ventricles in the transverse plane.

\subsection*{Clinical ECG recordings}
Twelve-lead ECGs of all induced ventricular tachycardia (VT) of patients with prior myocardial infarction who underwent radiofrequency catheter ablation (RFCA) for monomorphic VT at LUMC were reviewed. All patients provided informed consent and were treated according to the clinical protocol. Programmed electrical stimulation (PES) is routinely performed before RFCA to determine inducibility of the clinical/presumed clinical VT. All the patients underwent PES and ablation according to the standard clinical protocol, therefore no ethical approval was required. Ablation typically targets the substrate for scar-related reentry VT. After ablation PES is repeated to test for re-inducibility and evaluate morphology and cycle length of remaining VTs. The significance of non-clinical, fast VTs is unclear and these VTs are often not targeted by RFCA. PES consisted of three drive cycle lengths (600, 500 and 400 ms), one to three ventricular extrastimuli ($\geq$200 ms) and burst pacing (CL $\geq$200 ms) from at least two right ventricular (RV) sites and one LV site. A positive endpoint for stimulation is the induction of any sustained monomorphic VT lasting 30 s or requiring termination. ECG and intracardiac electrograms (EG) during PES were displayed and recorded simultaneously on a 48-channel acquisition system (Prucka CardioLab EP system, GE Healthcare, USA) for off-line analysis.

\bibliography{database.bib}

\section*{Author contributions statement}
N.V. conducted the experiments,  N.V., M.W. and A.P. analysed the results.  All authors reviewed the manuscript. 

\section*{Additional information}
 \textbf{Competing financial interests}: there are no competing interests.

\end{document}